\documentstyle[aps,preprint]{revtex}
\tighten

\makeatletter
\@addtoreset{equation}{section}
\makeatother

\begin{document}
\draft
%\twocolumn[
\hsize\textwidth\columnwidth\hsize\csname @twocolumnfalse\endcsname

\title{Quantum Statistical Mechanics of Ideal Gas Obeying Fractional
Exclusion Statistics: A Systematic Study\thanks{This paper is dedicated to Prof.
J. Zittartz on the occasion of his 60th birthday}}
\author{Gang Su$^{\dag}$ and Masuo Suzuki$^{\ddag}$}
\address{ Department of Applied Physics, Faculty of Science,
 Science University of Tokyo\\
1-3, Kagurazaka, Shinjuku-ku, Tokyo 162, Japan}
\date{22 January 1998}

\maketitle

\begin{abstract}
The quantum statistical mechanics of an ideal gas with a general
free-particle energy obeying fractional
exclusion statistics are systematically investigated in arbitrary
dimensions. The pressure relations,
the relation between pressure and internal
energy, the equation of state, as well as the thermodynamic properties are
thoroughly discussed. Some novel results are obtained.\\

\end{abstract}

\pacs{PACS numbers: 05.30.-d, 51.30, 71.10.+x}
%]

\section{Introduction}

In standard textbooks, quantum statistics usually refers to
the well-known Bose and Fermi ones. The former admits that a
state under permutation is symmetric, leading the maximum
occupation number of particles in one state to infinity, while
the latter requires that any state under permutation should be antisymmetric,
giving rise to the maximum occupation number in one state being one.
These two statistics are of fundamental significance for our understanding
of the real physical world nowadays. Indeed, a great number of physically basic
phenomena have been successfully explained within the framework of these
two celebrated statistics. Are they, however, adequate for us to describe
all macroscopic properties of the nature? The answer could be no. As a matter
of fact, after invention of these two statistics some people have been
considering how to generalize the quantum statistics to cover many
other cases.
A few such generalizations have therefore been done up to date.

One generalization is the so-called parastatistics (including
para-Bose and para-Fermi statistics)\cite{para}. Such a kind of statistics
is obtained in the following way: The expectation value of any
observable in the permuted state is required to be the same as in the
unpermuted state. It thus brings about two cases.
The para-Fermi statistics allows up to a finite number (greater than
one) of particles occupying one state, while the para-Bose statistics allows
that the wavefunction in one state vector can be antisymmetric with respect
to a finite number (greater than one) of particles but keeping the maximum
occupation number to infinity. This generalization has some implications
in quantum field theories, and is extensively studied in past decades.

Another generalization is named as the intermediate
statistics\cite{inter}.
This statistics is obtained by simply postulating that the maximum
occupation number (denoted by $\nu$) of particles in a single-particle
state is finite,
which naturally interpolates between Fermi ($\nu =1$) and Bose ($\nu
=\infty$) statistics. However, there is
neither a mathematical basis for the symmetry properties of wavefunctions
nor any generalized field quantization scheme available for intermediate
statistics. One of implications of this statistics can be attributed to
magnons in the Heisenberg model of magnets. As is well-known, the
maximum number of magnons in a system with a given number of spins is
finite, suggesting that magnons are not exactly bosons.

There are other generalizations (e.g., the q-deformed Fermi and Bose
statistics (see, e.g. Refs.\cite{su0}) motivated by the quantum
group theory, etc.). The most recent and
more physical generalization is the famous fractional exclusion
statistics (FES), first proposed by Haldane\cite{hald} and then realized
by Wu\cite{wu}. Inspired by studies in the fractional quantum Hall
systems and in one-dimensional (1d) exactly solvable models, and by
considering many particle systems with finite dimensional
Hilbert spaces where the dimensionality of the single-particle Hilbert
space depends {\it linearly} on the total number of particles,
Haldane\cite{hald} defined the exclusion statistics through
the changes of
the dimension of the single-particle space and the number of
particles. By
interpolating the statistical weights of Bose and Fermi statistics,
Wu\cite{wu} in his seminar paper derived the distribution function of
FES, which is usually called as Haldane-Wu distribution function in
literature. Recently it becomes aware that the physical realizations of FES
and associated generalized ideal gas can be reached in 1d integrable
models with long-range interactions of quantum fluids where the
particles are shown to obey FES, in the lowest Landau level where
anyons are shown to obey FES, and in the fractional quantum Hall
effect where the quasi-holes and the quasi-electrons obey FES\cite{wu2}.
Consequently, since the
appearance of Wu's remarkable paper a number of works on FES
have been done\cite{isakov,naya,murthy,rajag,sen,wuyu,fukui,fayya,iguchi}.
However, some basic issues, for instance, the fundamental
pressure inequalities
which are well known for Bose and Fermi statistics, a general relation
between the pressure and internal energy (Bernoulli equation),
as well as the thermodynamic properties, etc.
on quantum statistical mechanics (QSM)
of FES still need to be unambiguously addressed.
As FES has essential implications in
low-dimensional physical systems, a thorough discussion on it is
really interesting and necessary.

The outline of this paper is as follows.  In Sec. II the definition
and the
basic formulae of FES are briefly recalled. The pressure relations,
a relation between the pressure and internal energy, the
equation of state, as well as the thermodynamic properties are
presented in Secs. III-VI, respectively. Finally, a summary of results
is given.

\section{Fractional Exclusion Statistics}

Let us first briefly recall the definition and
some basic formulae of FES in this section,
which will be used in the subsequent discussions. By adopting a
state-counting definition, Wu\cite{wu} managed to write down the
statistical weight, say,  the number of quantum states of $N$
identical particles obeying FES occupying a group of $G$ states, as
\begin{equation}
W = \frac{[G + (N-1)(1-g)]!}{N! [G-gN -(1-g)]!},
\label{statweight}
\end{equation}
where the statistics parameter $g$ represents the number of states
that one particle can ``occupy'', or the ability of a particle to
exclude other particles in ``occupying single-particle state'', with
$g=0$ corresponding to usual bosons and $g=1$ fermions. (Hereafter we
only consider $ 0\leq g \leq 1$). By assuming
that an ideal gas, where every single-particle state of species $i$
has the same energy $\epsilon_{i}$, is a system with the total energy
taking a simple sum
\begin{equation}
E = \sum_{i} N_i \epsilon_i,
\label{energy-i}
\end{equation}
and using the standard argument of QSM, he obtained the grand partition
function
\begin{equation}
\Xi = \sum_{\{N_i\}} W(\{N_{i}\}) \exp \{-\beta \sum_{i}N_i
(\epsilon_i-\mu)\},
\label{parfunc-i}
\end{equation}
with $\beta = 1/k_{B}T$ the inverse temperaure ($k_{B}$ the Boltzmann
constant) and $\mu$ the chemical potential.
The most-probable distribution of $n_i$, the average
``occupation number'' defined by $N_i/G_i$, for identical particles
was shown\cite{wu} to be determined by
\begin{equation}
n_i = \frac{1}{\omega (e^{\beta (\epsilon_{i} -\mu)}) + g},
\label{n-i}
\end{equation}
where the function $\omega(\xi)$ satisfies the functional equation
\begin{equation}
\omega(\xi)^{g} [1 + \omega(\xi)]^{1-g} = \xi \equiv e^{\beta (\epsilon
-\mu)}.
\end{equation}
Eq.(\ref{n-i}) is the so-called Haldane-Wu distribution function.
The thermodynamic potential $\Omega$ and the total number of particles
$N$ can thus be obtained by
\begin{eqnarray}
& & \Omega = -P V = - k_{B}T \sum_{i} G_i \log (1 + \frac{1}{\omega
(\xi_{i})}),
\label{thermopoten-i} \\
& & N = \sum_{i} G_{i}\frac{1}{\omega (\xi_{i}) +g}.
\label{parti-i}
\end{eqnarray}
We note that the above equations are formulated in the representation
of states. By recognizing the fact that the grand partition function at
the most probable distribution can be factorizable, like the usual Bose
and Fermi cases, the summation over states in above equations can be
converted into that over momentum\cite{iguchi}. The average
occupation number in momentum space can be written down as
\begin{equation}
\langle n_{{\bf p}}\rangle = \frac{1}{\omega_{{\bf p}} +g }
\label{n-p}
\end{equation}
with the function $\omega_{{\bf p}}$ satisfying
\begin{equation}
\omega_{{\bf p}}^{g}(1+\omega_{{\bf p}})^{1-g} = e^{\beta
(\epsilon_{{\bf p}}-\mu)}.
\label{wp}
\end{equation}
The thermodynamic potential and the total number of particles can then
become as
\begin{eqnarray}
& & \Omega = -P V = - k_{B}T \sum_{{\bf p}} \log (1 +
\frac{1}{\omega_{{\bf p}}}),
\label{thermopoten-p} \\
& & N = \sum_{{\bf p}} \frac{1}{\omega_{{\bf p}} +g}.
\label{parti-p}
\end{eqnarray}
These are basic equations useful in the subsequent analyses.

\section{Pressure Relations}

It is well-known that for standard Bose and Fermi statistics
the corresponding pressures ($P_{B}$ and $P_{F}$) of the ideal gases
satisfy the following inequality\cite{suzuki}
\begin{equation}
P_{B} < P_{cl} < P_{F}
\label{PbPf-ineq}
\end{equation}
for the fixed density $\rho = N/V$, where $P_{cl}$ is the pressure
of an ideal gas obeying Maxwell-Boltzmann statistics (or classical
statistics henceforth), given by
\begin{equation}
 P_{cl} = k_{B}T \frac{N}{V} = \rho k_{B}T.
\label{P-cl}
\end{equation}
The physical meaning of the inequality (\ref{PbPf-ineq}) is as follows.
The quantum
effects in Bose and Fermi statistics introduce the so-called effective
``statistical interactions''. The particles obeying Fermi statistics
tend to expel other particles more than those obeying the classical
statistics (in the latter case no any interaction exists between
particles), while the particles obeying Bose statistics tend to
attract other particles,
yielding the possibility of condensation of bosons. In other words,
this inequality reflects the consequences of Pauli's exclusion principle
and Bose-Einstein condensation.

Now let us consider if there exists a similar inequality for FES.
From Eq.(\ref{thermopoten-p}) we know that the pressure $P$ is given
by
\begin{equation}
P =  \frac{k_{B}T}{V} \sum_{{\bf p}} \log (1 +
\frac{1}{\omega_{{\bf p}}}).
\label{pressure}
\end{equation}
As discussed in Ref.\cite{wu}, $\omega_{{\bf p}}$ is non-negative. By
utilizing the inequality $\log (1+x) >\frac{x}{1+x}$ for $x>0$, we
find
\begin{equation}
P > \frac{k_{B}T}{V} \sum_{{\bf p}} \frac{\langle n_{{\bf p}}\rangle}{
1 + (1-g)\langle n_{{\bf p}}\rangle}.
\label{p-inequ1}
\end{equation}
By Eq.(\ref{n-p}), we have
\begin{equation}
\langle n_{{\bf p}}\rangle \leq \frac{1}{g}.
\label{np-ineq}
\end{equation}
Incorporating inequalities (\ref{p-inequ1}) and (\ref{np-ineq})
 one can get
\begin{equation}
P > g P_{cl}
\label{p-inequ2}
\end{equation}
for fixed density $\rho$. This inequality states that the effective
interactions between particles obeying FES is more repulsive,
implying an exclusion property of FES. When $g=1$, Inequality
(\ref{p-inequ2}) recovers the second inequality of
(\ref{PbPf-ineq}). Therefore, the inequality
(\ref{p-inequ2}) can be regarded as a generalization of (\ref{PbPf-ineq}).

Now we consider the free-particle energy
$\epsilon_{{\bf p}}$ with the following general form:
\begin{equation}
\epsilon_{{\bf p}} = c_{0}\cdot\frac{{\bf p}^{\alpha}}{m}
\label{epsil-p}
\end{equation}
with constants $c_{0}$, $\alpha >0$. When $\alpha =2$ and $c_{0}=1/2$,
it reproduces the usual case of an ideal
gas. When $\alpha =1$, the 1d Calogero-Sutherland (CS) model \cite{murthy}
falls into this class with a properly scaled value of $c_{0}$. However, the
following analyses hold for arbitrary $\alpha>0$. The density of states
(DOS) in $d$ dimensions can thus be obtained by
\begin{eqnarray}
& & {\cal D}(\epsilon) = A(\alpha, d) \epsilon^{\frac{d}{\alpha}-1}
\label{dos}\\
& & A(\alpha, d) = \frac{S_{d} (m/c_{0})^{d/\alpha}}{\alpha (2 \pi
\hbar)^{d}}, ~~~S_{d} = \frac{2 \pi^{d/2}}{\Gamma(\frac{d}{2})},
\label{A}
\end{eqnarray}
where we have transformed the summation over momentum into an integral:
$(1/V) \sum_{{\bf p}} \rightarrow \frac{S_{d}}{(2\pi \hbar)^{d}} \int
p^{d-1} dp$.

When $\alpha =d$, the DOS is constant: ${\cal D}(\epsilon) = A(d, d)$.
The cases, like an ideal gas with $\alpha =2$ in two dimensions
discussed in Refs.\cite{wu,wu2}, as well as the 1d CS model
considered in Ref.\cite{murthy}, belong to this category. Under the
condition of ${\cal D}(\epsilon)$ being constant, by noticing
$(1/V) \sum_{{\bf p}} \rightarrow \int {\cal D}(\epsilon) d\epsilon$
and $\rho = A(d, d)\int_{0}^{\infty}\frac{d\epsilon}{\omega (\epsilon) +g}$
one can get
\begin{equation}
\mu = k_{B}T \log (e^{\frac{\beta \rho}{A(d, d)}}-1) + (g-1)
\frac{\rho}{A(d, d)}.
\label{mu}
\end{equation}
Substituting (\ref{mu}) into $\frac{\partial P}{\partial g}$
 and working out the integral, we find
\begin{equation}
\frac{\partial P}{\partial g} = \frac{\rho^2}{2 A(d, d)} >0,
\end{equation}
implying that $P$ is a monotonically increasing function of $g$ in this
special case. In addition, Eq.(\ref{mu})
implies
\begin{equation}
\mu = g \mu_{F} + (1-g) \mu_{B}
\label{mu-fb}
\end{equation}
with $\mu_{F} = \mu (g=1)$ and $\mu_{B} = \mu (g=0)$. Considering
$\frac{\partial P}{\partial \mu} = \rho$, it can be seen that
\begin{equation}
P = g P_{F} + (1-g) P_{B}
\label{P-fb}
\end{equation}
for fixed $\rho$, which is also in agreement with (\ref{p-inequ2}).
Note that equation
(\ref{P-fb}) first appeared as a comment by Suzuki\cite{wu2} for
$\alpha =2$ and $d=2$.
To summarize, we have the following theorem\cite{note1}:

{\bf Theorem}: {\it For an ideal gas obeying FES with constant density
of states in d dimensions, $\mu = g \mu_{F} + (1-g) \mu_{B}$, and
$P = g P_{F} + (1-g) P_{B}$ for $0\leq g \leq 1$.}

A direct corollary of this theorem is: $\Xi = \Xi_{F}^{g}
\Xi_{B}^{1-g}$, suggesting that in this special case the ideal gas
with FES can be regarded as composites of usual fermions and bosons.
Unlike the assertion in Ref.\cite{iguchi} where some errors remain in
the proof, we should stress here that this statement could not hold
true for the case where DOS is not constant.

\section{The Bernoulli Equation}

Now let us discuss the general relation between the pressure and the
internal energy under assumption of the free-particle energy with the
form of (\ref{epsil-p}). The internal energy $E$ can be expressed as
\begin{equation}
E = \sum_{{\bf p}} \frac{\epsilon_{{\bf p}}}{\omega_{\bf p} +g}
= V \int_{0}^{\infty} \frac{{\cal D}(\epsilon) \epsilon d\epsilon}{
\omega (\epsilon) +g}.
\label{internal-E}
\end{equation}
By noting
$\frac{d}{d\epsilon} \log [ 1+ \frac{1}{\omega(\epsilon)}] = -
\frac{\beta}{\omega(\epsilon)+g}$, and integrating
Eq.(\ref{internal-E}) by part, we obtain
\begin{eqnarray}
 \frac{E}{V} & = &- \frac{1}{\beta} \log [ 1+ \frac{1}{\omega(\epsilon)}]
{\cal D}(\epsilon) \epsilon|_{0}^{\infty} + \frac{d}{\alpha \beta}
\int_{0}^{\infty} \log [ 1+ \frac{1}{\omega(\epsilon)}]
{\cal D}(\epsilon) d\epsilon  \nonumber \\
 & = &\frac{d}{\alpha} P,
\nonumber
\end{eqnarray}
giving rise to the Bernoulli equation
\begin{equation}
PV = \frac{\alpha}{d}E.
\label{bernoulli}
\end{equation}
It can be noted that Eq.(\ref{bernoulli}) is $g$-independent, as
it should be. For $\alpha =d$, $PV=E$, which is fit for the case of
1d CS model and the case of the usual ideal gas with $\alpha =2$ in two
dimensions. We would like to point out here that Eq.(\ref{bernoulli})
was ever derived for Bose and Fermi ideal gases by
Suzuki\cite{suzuki} for $d=3$. Now, we find that the Bernoulli
equation can be extended to cover an ideal gas with FES.

In addition, using Eqs.(\ref{pressure}), (\ref{bernoulli}) and
(\ref{wp}), one can find the pressure $P(T)$ can be determined by
the following functional
\begin{equation}
P(T) = \rho \mu (T) + C_{1} T + \frac{d}{\alpha} T \int^{T}
\frac{P(x)}{x^2} dx,
\end{equation}
where $C_{1}$ is a temperature-independent constant that can be
determined by known conditions. This equation can be solved numerically.

\section{Equation of State}

In this section, we shall discuss the equation of state, namely, the
Virial expansion for an ideal gas obeying FES. Define
\begin{equation}
z = e^{\beta \mu}
\end{equation}
as the fugacity. In parallel to Bose and Fermi statistics\cite{huang},
we can expand $\frac{P}{k_{B}T}$ and $N/V$ in powers of $z$, and get
\begin{equation}
\frac{PV}{Nk_{B}T} = \frac{\sum_{l=1}^{\infty} b_{l} z^{l}}{
\sum_{l=1}^{\infty} l b_{l} z^{l}}
\label{z-expan}
\end{equation}
with $b_{l}$ the expansion coefficients. The Virial expansion gives
\begin{equation}
\frac{PV}{Nk_{B}T} = \sum_{l=1}^{\infty} a_{l} (\lambda^{d}\rho)^{l-1}
\label{virial}
\end{equation}
with $\lambda$ to be determined later, and $\lambda^{d} \rho<1$.
By expanding $\rho = \frac{N}{V}$ in the right-hand side (r.h.s.) of
(\ref{virial}) in powers of $z$, and comparing the
coefficients of Eqs.(\ref{z-expan}) and (\ref{virial}), one can obtain
the virial coefficients\cite{huang},
\begin{eqnarray}
& & a_{1} = b_{1} = 1, ~~a_{2} = - b_{2}, ~~a_{3} = 4 b_{2}^2 -2 b_{3},
\nonumber \\
& & a_{4} = -20 b_{2}^{3} + 18 b_{2}b_{3} - 3b_{4}, ~~\cdots.
\end{eqnarray}
Now, let us calculate the coefficients $b_{l}$. By expanding the
logarithmic function in $P$ in powers of $ze^{-\beta \epsilon_{{\bf p}}}$,
we have
\begin{equation}
\log (1 + \frac{1}{\omega_{{\bf p}}}) = \sum_{l=1}^{\infty} c_{l}
(e^{-\beta \epsilon_{{\bf p}}}z)^{l}
\label{log-expan}
\end{equation}
for $e^{\beta (\epsilon_{{\bf p}}-\mu)} >1$. Since this is a standard
Taylor expansion, we can obtain the coefficients as
\begin{eqnarray}
& & c_{1} =1, ~~c_{2} = -\frac{1}{2!}(2g-1), ~~c_{3} = \frac{1}{3!}
(3g-1)(3g-2), \nonumber \\
& & c_{4} = -\frac{1}{4!} (4g-1)(4g-2)(4g-3), ~~\cdots, ~~c_{l}=
\frac{(-1)^{l-1}}{l!} \prod_{m=2}^{l} [lg -(m-1)] ~for~ l\geq 2.
\end{eqnarray}
Substituting (\ref{log-expan}) into (\ref{pressure}) yields
\begin{eqnarray}
& & \frac{P}{k_{B}T} = \int_{0}^{\infty}\sum_{l=1}^{\infty} c_{l} z^{l}
e^{-\beta \epsilon l} {\cal D}(\epsilon) d\epsilon \nonumber \\
& & = \sum_{l=1}^{\infty} \frac{c_{l}A(\alpha, d)}{(\beta
l)^{\frac{d}{\alpha}}} \Gamma(\frac{d}{\alpha}) z^{l}
= \sum_{l=1}^{\infty} \frac{b_{l}}{\lambda^{d}}z^{l}
\label{expansion}
\end{eqnarray}
with
\begin{equation}
b_{l} = \frac{c_{l}}{l^{\frac{d}{\alpha}}}, ~~
\lambda ^{d} = \frac{\beta^{\frac{d}{\alpha}}}{A(\alpha, d) \Gamma
(\frac{d}{\alpha})}.
\end{equation}
 Consequently, the Virial expansion coefficients are given by
\begin{eqnarray}
& & a_{1} =1, \nonumber \\
& & a_{2} = \frac{2g-1}{2^{\frac{d}{\alpha}+1}}, \nonumber \\
& & a_{3} = \frac{(2g-1)^2}{4^{\frac{d}{\alpha}}} -
\frac{(3g-1)(3g-2)}{3^{\frac{d}{\alpha}+1}}, \nonumber \\
& & a_{4} = 20 (\frac{2g-1}{2^{\frac{d}{\alpha}+1}})^{3} -
\frac{3(2g-1)(3g-1)(3g-2)}{2 \cdot 6^{\frac{d}{\alpha}}}
+ \frac{(4g-1)(4g-2)(4g-3)}{2 \cdot 4^{\frac{d}{\alpha}+1}}, ~~\cdots,
\end{eqnarray}
using (5.4). We note that the similar results are obtained for
$\alpha =2$\cite{iguch2}, but our results are more general and include
the results obtained in Ref.\cite{iguch2} as a special case.
The equation of state can then have the form
\begin{equation}
\frac{PV}{Nk_{B}T} = 1 + a_{2}{\tilde \rho} + a_{3} {\tilde \rho}^{2}
+ a_{4} {\tilde \rho}^{3} + \cdots
\label{state}
\end{equation}
with ${\tilde \rho} = {\lambda}^{d}\rho$. In the limit of ${\tilde \rho}
\rightarrow 0$, we find that the effective ``statistical interaction''
is repulsive for $g \geq \frac12$ and attractive for $g<\frac12$. We
notice that there is a similar discussion on this point in
Ref.\cite{iguch2}, where he finds that no effective statistical
interaction exists for semions (with $g=\frac12$). This is incorrect,
because for $g=\frac12$, $a_{2}=0$, but $a_{3}>0$, suggesting the
effective statistical interactions between semions are
repulsive.
 For ${\tilde \rho}$ ($<1$) not so small, to
discuss the property of effective interactions between particles
obeying FES we have to take the $a_{3}$ term in above equation into
account, which gives if
\begin{equation}
g \geq \frac{-\frac{1}{4^{\frac{d}{\alpha}-1}} +
\frac{1}{3^{\frac{d}{\alpha}-1}} +
\frac{1}{2^{\frac{d}{\alpha}}{\tilde \rho}} -
\sqrt{(\frac{1}{2^{\frac{d}{\alpha}}{\tilde \rho}})^2 +
(\frac{1}{3^{\frac{d}{\alpha}}})^2
- \frac{1}{
3^{\frac{d}{\alpha}+1}\cdot
4^{\frac{d}{\alpha}-1}} }}{2 (\frac{1}{3^{\frac{d}{\alpha}-1}}
-\frac{1}{4^{\frac{d}{\alpha}-1}})},
\label{g>1/2}
\end{equation}
the effective statistical interaction is repulsive.  The r.h.s. of
(\ref{g>1/2}) is smaller than $1/2$, while ${\tilde \rho}
\rightarrow 0$, it gives $1/2$. This fact implies that for the density
$\rho$ fixed there must be a critical value $g_{c}$ at which no effective
statistical interactions exist between particles with FES. In other
words, the effective interactions between particles would be attractive
for $g<g_{c}$ and repulsive for $g>g_{c}$. Here $g_{c}$ can be
self-consistently determined from the following equation
\begin{equation}
\rho = \int_{0}^{\infty} d\epsilon {\cal D}(\epsilon) \log [1 +
\frac{1}{\omega(\epsilon, g_{c})}],
\end{equation}
where the chemical potential $\mu$ can be obtained through $\rho =
\int_{0}^{\infty} d\epsilon \frac{{\cal D}(\epsilon)}{\omega(\epsilon,
g_{c}, \mu) + g_{c}}$, and the function $\omega$ is determined by
(\ref{wp}). We observe that $g_{c}$ depends on the density
$\rho$. The similar result has been reached in the case of
constant DOS and for $\alpha =2$ a few years ago\cite{suzuki}.

\section{Thermodynamic Properties}

From (\ref{n-p}), one may find that the average occupation
number $\langle n_{{\bf p}} \rangle$ is bounded by $1/g$, implying
that $\langle n_{{\bf p}} \rangle$ can not be {\it
macroscopically} large for any momentum except $g=0$, and thereby leading
to the fact that no any condensation phenomenon occurs in the ideal
gas obeying FES for $0<g\leq 1$. For the free-particle energy given
by (\ref{epsil-p}), the  Fermi energy  for FES is
\begin{equation}
\epsilon_{F} = [\frac{\rho g d}{\alpha A(\alpha,
d)}]^{\frac{\alpha}{d}}.
\end{equation}

To discuss the thermodynamic properties, we need to perform
the Sommerfeld expansion. As a result, the expansion gives
\begin{eqnarray}
I &=& \int_{0}^{\infty} g(\epsilon) f(\epsilon) d\epsilon \nonumber \\
  &=& \frac{G(\mu)}{\omega(0)+g} - \frac{G'(\mu) \mu}{\omega(0)+g}
  + \frac{G''(\mu) \mu^2/2}{\omega(0)+g}+ \frac{G'(\mu)-\mu G''(\mu)}{\beta}
  \log [1 + \frac{1}{\omega(0)}] \nonumber \\
  &+& \frac{G''(\mu)}{2 \beta^2}[\omega(0)+g] \log^2 [1 + \frac{1}{\omega(0)}]
  + \frac{G''(\mu)}{2 \beta^2} \phi [\log (1 + \frac{1}{\omega(0)})],
\label{sommerfeld}
\end{eqnarray}
where $g(x)$ is any function of $x$, and
\begin{eqnarray}
f(x) = \frac{1}{\omega(x) +g}, \\
G(x) = \int_{0}^{x} g(t) dt, \\
\phi (x) = \int_{0}^{x} \frac{t^2 e^{t}}{(e^{t}-1)^2} dt,
\end{eqnarray}
with  the notation $G'(x) =dG(x)/dx$ and $G''(x) = d^2G(x)/dx^2$.
$\omega(0)$ is determined by the functional
\begin{equation}
\omega(0)^g [1 + \omega(0)]^{1-g} = e^{-\beta \mu}.
\label{omega0}
\end{equation}
By using this expansion, the thermodynamic quantities can be
obtained. For instance, the density $\rho =N/V$ can be gained
by setting $g(\epsilon) = {\cal
D}(\epsilon)$. For fixed $\rho$, $\mu = \mu(g,T)$ can in turn be determined.
By setting $g(\epsilon) = \epsilon {\cal D}(\epsilon)$, one can get the
internal energy $E$ as
\begin{eqnarray}
\frac{E}{V} &=& \frac{A(\alpha, d)\mu^{\frac{d}{\alpha}+1}}{\omega(0)+g}
\cdot \frac{d(d-2\alpha)}{2\alpha (\alpha +d)} + \frac{A(\alpha,d)
\mu^{\frac{d}{\alpha}}}{\beta} (1-\frac{d}{\alpha})
\log [1 + \frac{1}{\omega(0)}] \nonumber \\
&+& \frac{A(\alpha, d)}{2 \beta^2}
\frac{d}{\alpha}\mu^{\frac{d}{\alpha}-1}\{ [\omega(0)+g]
\log^2 [1 + \frac{1}{\omega(0)}] + \phi [\log (1 +
\frac{1}{\omega(0)})]\}.
\label{energy}
\end{eqnarray}
The specific heat per volume can thus be obtained via
$c_{V}(g,T) = (\frac{\partial E(T)/V}{\partial T})_{N,V}$. The
temperature-dependence of $c_{V}(g,T)$ can be numerically
determined by Eqs. (\ref{omega0}), (\ref{energy}) and
$\mu = \mu(g,T)$ self-consistently  in general.
At very low temperature, $T\rightarrow
0$, $\mu \approx \mu_{0} \equiv \mu(T=0)$, $\omega(0) \sim e^{-\beta
\mu_{0}}$, for $0<g\leq1$ and $g\neq 0$
one may find analytically
\begin{equation}
c_{V}(g,T) \sim {\tilde c}_{0} T, ~~~{\tilde c}_{0} \approx
\frac{\pi^2 d}{3\alpha} k_{B}^2
A(\alpha, d) \mu_{0}^{\frac{d}{\alpha}-1},
\end{equation}
where we have used $\phi (\infty) = \pi^2/3$. It suggests that at very low
temperatures the specific heat depends linearly on temperature and
goes to zero at T=0, like an Fermi ideal gas.

For $d=\alpha$, i.e., in the case of constant DOS, we can get a closed
form for $E/V$:
\begin{equation}
\frac{E}{V} = \frac{\rho^2}{2 A(d,d)} (g + \frac{1}{
e^{\frac{\rho}{A(d,d) k_{B} T}}-1}) + \frac{A(d,d) k_{B}^2
T^2}{2}\phi (\frac{\rho}{A(d,d) k_{B} T}).
\end{equation}
The specific heat $c_{V}$ for fixed $\rho$ is thus
\begin{eqnarray}
c_{V}(T) = A(d,d) k_{B}^2 T \phi(\frac{\rho}{A(d,d)
k_{B} T}).
\label{capacity}
\end{eqnarray}
In this case, the specific heat is independent of $g$ for fixed $\rho$,
and goes to the
classical value ($k_{B}\rho$) at $T \rightarrow \infty$ as it should be.
To see qualitatively the temperature-dependence of the specific heat,
we depict Eq. (\ref{capacity}) as an example, as shown in Fig. 1.

The Sommerfeld expansion (\ref{sommerfeld}) could be used to
investigate the magnetic susceptibility of an ideal gas obeying FES to
see if the Pauli paramagnetism still sustains in the system apart from
$g=1$. This work is left to study in future.

\section{Summary}

In this paper we investigate systematically the quantum statistical
mechanics of an ideal gas, with a general free-particle energy,
obeying FES. Almost all basic properties of this
simple system are studied. Some results appear for the first time, and
some previously obtained results are extended to cover more general cases.
A general
inequality among the pressure for arbitrary statistics parameter
$g$ and for the classical and Bose
statistics is obtained for the fixed density, which could gain
somewhat insight into
the effective statistical interactions between particles with FES.
 For the density of states being constant, we present
a theorem for the chemical potential and the pressure, which corrects
some erroneous statements asserted by other people. A general relation
between pressure and internal energy, say, the Bernoulli equation,
is proved to hold for this system. The Virial expansion for this system
is studied for a general case, and some wrong assertions by other
people are also corrected. A critical value for $g$, below which the effective
statistical interactions between particles are found to be attractive
and above which repulsive, is predicted, and the equations used to
determine $g_{c}$ numerically are given.
Finally, the Sommerfeld expansion is done, which allows
to discuss the thermodynamic properties of the system. It is shown
that at very low temperatures the specific heat depends linearly on
temperature and vanishes at zero temperature. For the case with a
constant density of states, a closed form for the specific heat is
presented. We expect that the results obtained in this paper
would be useful to further understand
the characteristics of the particles (e.g., anyons, semions, or
quasi-particles in the fractional quantum Hall effect and in other
low-dimensional electron or spin systems) with FES.

Moreover, there are a lot of interesting questions remaining open. For
instance, how to construct a theory on QSM when weak interactions exist
between particles obeying FES, or a generalized Landau Fermi liquid
theory using FES, how to make the results obtained in
this paper experimentally measurable, and if there also exists the
de Haas-van Alphen effect in the system with FES, and so forth, deserve
to discuss and to explore. Therefore, many fascinating but difficult problems
remain in this new field.

\acknowledgments

One of authors (GS) expresses his faithful acknowledgements to
Prof. J. Zittartz for spending very pleasant and profitable time in
his group, and cordially wishes him many more happy years in scientific
activities. He
is grateful to the Department of Applied Physics,
Science University of Tokyo, for the warm hospitality, and
to the Nishina
foundation for support. This work has also been supported
in part by the CREST
(Core Research for Evolutional Science and Technology) of the Japan
Science and Technology Corporation (JST).

\vspace{2.0cm}

{\bf Figure Caption} \\

Fig. 1 The temperature-dependence of the specific heat for fixed
desity with constant DOS.


\begin{references}
\bibitem{note1}We note that a similar theorem was stated in
Ref.\cite{iguchi} for a general case. However, the proof presented
by this author contains some serious errors, and thereby his statement
is not justified, and could be incorrect.
\end{references}
\end{document}